\documentclass{IETpaper}
\usepackage[noadjust]{cite}
\usepackage{graphicx}
\usepackage{epstopdf}
\usepackage[cmex10]{amsmath}
\usepackage{amsfonts}
\usepackage{amssymb,amsthm}
\usepackage{array,multirow}
\usepackage[font=footnotesize]{subfig}
\usepackage{setspace, color,graphicx,xcolor,url, mathrsfs,mathtools}
\usepackage{fixltx2e}

\usepackage[all=normal,floats,paragraphs,charwidths,wordspacing,
tracking,mathspacing,bibliography,bibbreaks,bibnotes,leading]{savetrees}
\DeclareMathOperator*{\argmin}{\arg \! \mbox{   } \min}

\theoremstyle{definition}
\newtheorem{defn}{\bf Definition}

\theoremstyle{theorem}
\newtheorem{thm}{\bf Theorem}

\theoremstyle{remark}
\newtheorem{remk}{\bf Remark}

\begin{document}


\title{Constrained Least Squares, SDP, and QCQP Perspectives on Joint Biconvex Radar Receiver and Waveform design}

\author{Pawan Setlur$^{\ast \dag}$, Sean M. O'Rourke$^{\ast}$, Muralidhar Rangaswamy$^{\ast}$, }

\institution{$^{*}$US Air Force Research Laboratory,
Sensors Directorate, WPAFB, OH 45433,
$^{\dag}$Wright State Research Institute,
Beavercreek, OH 45431}

\maketitle


\begin{abstract}
Joint radar receive filter and waveform design is non-convex, but is individually convex for a fixed receiver filter while optimizing the waveform, and vice versa. Such classes of problems are frequently encountered in optimization, and are referred to biconvex programs. Alternating minimization (AM) is perhaps the most popular, effective, and simplest algorithm that can deal with bi-convexity. In this paper we consider new perspectives on this problem via older, well established problems in the optimization literature. It is shown here specifically that the radar waveform optimization may be cast as constrained least squares, semi-definite programs (SDP), and quadratically constrained quadratic programs (QCQP). The bi-convex constraint introduces sets which vary for each iteration in the alternating minimization. We prove convergence of alternating minimization for biconvex problems with biconvex constraints by showing the equivalence of this to a biconvex problem with constrained Cartesian product convex sets but for convex hulls of small diameter.
\end{abstract}

\section{Introduction}
We address waveform design in radar space time adaptive processing (STAP) \cite{klemm2002,ward1994,guerci2003,Brennan1973,SetlurTSP2016,
Setlurrangareport2014,ORourkeAFRLTR2016}. An air-borne radar is assumed with an array of sensor elements observing a moving target on the ground. We will assume that the waveform design and scheduling are performed over one CPI rather than on an individual pulse repetition interval (PRI).

In line with traditional STAP, the optimization is cast as an minimum variance distortion-less response (MVDR) type optimization \cite{Setlurrangareport2014,SetlurTSP2016,ORourkeAFRLTR2016}. Classical Radar STAP is computationally expensive but waveform-adaptive STAP increases the complexity by several orders of magnitude. Therefore, the benefits of waveform design in STAP come at the expense of increased computational complexity. 

It was shown in \cite{Setlurrangareport2014,SetlurTSP2016} that the optimization problem is biconvex \cite{ORourkeAFRLTR2016}, therefore alternating minimization was used. A convex relaxation approach was pursued in \cite{ORourkeAFRLTR2016}. 

{\bf Contributions:} Here, we also employ alternating minimization, but our contributions in this paper are to demonstrate the relationships of this radar optimization problem to several well established convex programming constructs in the literature. In particular, we show that in the waveform design stage, the problem may be cast as an SDP, a QCQP with a single constraint, and a constrained least squares problem on hyper-ellipses.  The bi-convexity in the joint radar receive filter and waveform optimization introduces varying constraint sets, in particular, iteration varying constraint sets. That is, at each iteration, the constraint sets differ from the previous iteration. Our other contribution is to demonstrate the convergence of the alternating minimization to biconvex problems, and develop the necessary conditions when this occurs. As a first step, we consider certain assumptions where our proofs are valid. Specifically, those regions where the convex hull of the iterates of the alternating minimization is small.

\section{Preliminaries}
We consider the definition of a bi-convex problem and then delve into the radar specifics. 
\subsection{Bi-convex programming}
Consider an objective function $f(\mathbf{x},\mathbf{y}):\mathbb{F}^{N}\times \mathbb{F}^{M}\rightarrow \mathbb{R}$, with two vector parameters, $\mathbf{x},\mathbf{y}$, and where the field $\mathbb{F}=\mathbb{R} \mbox{ or } \mathbb{C}$.
\begin{defn}[Bi-convex] \label{def1} An optimization problem,
\begin{align} \label{biconvex1}
\min \limits_{\mathbf{x},\mathbf{y}} \,\,\,\,\,\,\,\, &f(\mathbf{x},\mathbf{y})\nonumber\\
\mbox{s. t. }\,\,\,\,\,\,\,\,&g_i(\mathbf{x},\mathbf{y})\leq 0,i=1,2,\ldots, \\\
\,\,\,\,\,\,\,\,&h_j(\mathbf{x},\mathbf{y}) =0,j=1,2,\ldots,\nonumber
\end{align}
is bi-convex if and only if, for some $\tilde{\mathbf{x}} \times \mathbf{y} \in \mbox{dom}(f(\cdot,\cdot))$ and for some $\mathbf{x} \times \tilde{\mathbf{y}} \in \mbox{dom}(f(\cdot,\cdot))$,
\begin{enumerate}
\item the functions $f(\mathbf{x},\;\;\mathbf{y}=\tilde{\mathbf{y}})$ and $f(\mathbf{x}=\tilde{\mathbf{x}},\;\;\mathbf{y})$,  are convex in $\mathbf{x},\mathbf{y}$ respectively, and
\item if the inequality constraint functions $g_i(\mathbf{x},\;\;\mathbf{y}=\tilde{\mathbf{y}})\leq 0$ and $g_i(\mathbf{x}=\tilde{\mathbf{x}},\;\;\mathbf{y})\leq 0$ are convex in $\mathbf{x},\mathbf{y}$ respectively, for all $i=1,2,\ldots$, and 
\item if the equality constraints $h_j(\mathbf{x},\;\;\mathbf{y}=\tilde{\mathbf{y}})=0$ and $h_j(\mathbf{x}=\tilde{\mathbf{x}},\;\;\mathbf{y})=0$ are convex in $\mathbf{x},\mathbf{y}$ respectively, for all $j=1,2,\ldots$.
\end{enumerate}
\end{defn}
We note that bi-convex problems are special cases of non-convex problems and, in many cases, are generally easier to solve than a general non-convex optimization problem. In essence, bi-convexity is a milder form of non-convexity.

The alternating minimization (AM) algorithm operates by initializing $\mathbf{x}=\mathbf{x_0}\in\mathcal{X}_0, \; \mathbf{y} = \mathbf{y_0} \in \mathcal{Y}_o$, and then at each iteration solves the reduced dimension optimizations
\begin{subequations} \label{ambiconvex}
\begin{align} 
\mathbf{y}_k=\argmin \limits_{\mathbf{y}} \,\,\,\,\,\,\,\,  &f(\mathbf{x}_k,\mathbf{y})\nonumber\\ 
\mbox{s. t. }\,\,\,\,\,\,\,\, &g_i(\mathbf{x}_k,\mathbf{y})\leq 0,i=1,2,\ldots, \label{ambiconvex1} \\
\,\,\,\,\,\,\,\, &h_j(\mathbf{x}_k,\mathbf{y}) =0,j=1,2,\ldots,\nonumber\\
\mathbf{x}_{k+1}=\argmin \limits_{\mathbf{x}} \,\,\,\,\,\,\,\,  &f(\mathbf{x},\mathbf{y}_K)\nonumber\\
\mbox{s. t. }\,\,\,\,\,\,\,\, &g_i(\mathbf{x},\mathbf{y}_k)\leq 0,i=1,2,\ldots, \label{ambiconvex2} \\
\,\,\,\,\,\,\,\, &h_j(\mathbf{x},\mathbf{y}_k) =0,j=1,2,\ldots,\nonumber
\end{align}
\end{subequations}
for some $k=0,1,2,\ldots$. This simple, yet powerful idea is long-standing, and could possibly be attributed to John Von Neumann's work in the 1930s. 
Nonetheless, alternating minimization and its variants (e.g. block co-ordinate descent, or Gauss-Seidel etc.) are emerging as front runners in many deep learning and big data optimization challenges. 

The AM algorithm by construction affords an important property -- that is, a monotonic cost function decrease at each iteration, i.e.
\begin{align*}
&f(\mathbf{x_0},\mathbf{y_0}) \geq f(\mathbf{x_1},\mathbf{y_0}) \geq f(\mathbf{x_1},\mathbf{y_1}) \geq \\
&\cdots f(\mathbf{x_k},\mathbf{y_{k-1}})\geq f(\mathbf{x_k},\mathbf{y_k}) \cdots.
\end{align*}

\subsection{Radar model}
Consider an airborne radar STAP detection problem, with $M$ sensors, $N$ fast time samples, and $L$ waveform repetitions. Assume a hypothesized target at  a particular range gate, and Doppler $f_d$ and at spatial co-ordinates at $\theta_t,\phi_t$. The detection problem is cast as
\begin{align*} 
&\mathcal{H}_0: \,\,\,\, \mathbf{y_r}= \mathbf{y_u} \\
&\mathcal{H}_1: \,\,\,\, \mathbf{y_r}=\mathbf{y_t}+\mathbf{y_u}
\end{align*}
where $\mathbf{y_r}$ is the radar measurement at the chosen range gate. The vectors $\mathbf{y_t}, \mathbf{y_u}$ are the hypothesized deterministic target response and the stochastic clutter-plus-interference-plus-noise measurement, respectively, and are assumed statistically independent from each other. If the waveform is defined as $\mathbf{s}\in\mathbb{C}^{N}$, then $\mathbf{y_t}=\rho_t\mathbf{v}(f_d)\otimes\mathbf{s}\otimes\mathbf{a}(\theta_t,\phi_t)$, where $\rho_t$ is a complex scalar target response, and $\mathbf{v}(f_d),\mathbf{a}(\theta_t,\phi_t)$ are the Doppler and spatial steering vectors, respectively, and $\otimes$ is the Kronecker product. 

Now let the radar receiver use a weight vector $\mathbf{w}\in\mathbb{C}^{MNL}$ to process the radar returns.  The covariance matrix of $\mathbf{y_u}$ is $\mathbf{R}_{\bf u}(\mathbf{s})=\mathbf{R}_{\bf c}(\mathbf{s})+\mathbf{R}_{\bf n} +\mathbf{R}_{\bf i}$ i.e. the clutter plus interference plus noise covariance matrix, and as such is a function of the waveform $\mathbf{s}$. This is because the clutter is dependent on the transmitted waveform, whereas the noise and interference are not. More details on the radar STAP model for joint receiver and waveform design can be seen in \cite{Setlurrangareport2014},\cite{SetlurTSP2016}.

\section{Joint Radar Receiver and Waveform Design}
The radar return at the considered range gate is processed by a filter characterized by a weight vector, $\mathbf{w}$, whose output is given by $\mathbf{w}^H\mathbf{y}_{\bf r}$. Since the vector $\mathbf{s} \in\mathbb{C}^{N}$ prominently figures in the steering vectors, the objective is to jointly obtain the desired weight vector $\mathbf{w}$ and waveform vector $\mathbf{s}$. It is desired that the weight vector will minimize the output power, $\mathbb{E}\{|\mathbf{w}^H\mathbf{y_u}|^2\}=\mathbf{w}^H\mathbf{R_u}( \mathbf{s})\mathbf{w}$. Mathematically, we may formulate this problem as:
\begin{align}
\min_{\mathbf{w},\mathbf{s}}\;\;\;\;\; &\mathbf{w}^H\mathbf{R_u}( \mathbf{s})\mathbf{w}\label{eq1}  \nonumber\\
\mbox{ s. t } \;\;\;\;\;&\mathbf{w}^H(\mathbf{v}(f_d)\otimes\mathbf{s}\otimes\mathbf{a}(\theta_t,\phi_t))=\kappa  \\
\;\;\;\;\;\; & \mathbf{s}^H \mathbf{s}\leq P_o \nonumber \nonumber
\end{align}
In \eqref{eq1}, the first constraint is the well known Capon constraint where, typically, $\kappa=1$. An energy constraint, enforced via the second constraint, addresses hardware limitations in the radar. The optimization in \eqref{eq1} was shown to be  bi-convex according to Definition \ref{def1} in \cite{Setlurrangareport2014,SetlurTSP2016}.

If AM is used to solve \eqref{eq1}, we have, at the $k$-th iteration \cite{Setlurrangareport2014,SetlurTSP2016}, a weight vector and waveform update given by
\begin{subequations} \label{weightcomp}
\begin{align}
\mathbf{w_k}&=\frac{\kappa \mathbf{R}_{\bf u}^{-1}(\mathbf{s_{k-1}}) \mathbf{G} \mathbf{s_{k-1}}}{\mathbf{s_{k-1}}^H\mathbf{G}^H\mathbf{R}_{\bf u}^{-1} (\mathbf{s_{k-1}}) \mathbf{G} \mathbf{s_{k-1}}}  \label{weightcomp1}\\
\mathbf{s_{k}}&=\frac{\kappa \mathbf{F}^{-1}\mathbf{G}^H\mathbf{w_k} }{\mathbf{w_K}^H\mathbf{G} \mathbf{F}^{-1}\mathbf{G}^H\mathbf{w_k}} \label{weightcomp2}
\end{align}
\end{subequations}
where, $\mathbf{v}(f_d)\otimes s \otimes \mathbf{a}(\theta_t,\phi_t)=\mathbf{Gs}$, and $\mathbf{F}=\left( \sum\limits_{q=1}^Q \mathbf{Z_q}( \mathbf{w_k})+\lambda\mathbf{I} \right) $. The matrix $\sum\limits_{q=1}^Q \mathbf{Z_q}( \mathbf{w_k})$ is related to the sum of the covariance of the $Q$ clutter patches.  Additional details on this relationship are unnecessary here, since it digresses from the main focus; however, these may be found in \cite{Setlurrangareport2014,SetlurTSP2016}. We also have $ \lambda $ as a Lagrange parameter, where
\begin{align} \label{rootsol}
&\lambda=\max [0,\lambda^{\ast}] \\
&\lambda^{\ast} \mbox{ solves } \lambda^{\ast}\left( \kappa^2\mathbf{w}^H\mathbf{G}\mathbf{F}^{-2}\mathbf{G}^H\mathbf{w}-P_o( \mathbf{w}^H\mathbf{G}\mathbf{F}^{-1}\mathbf{G}^H\mathbf{w})^2\right)=0 \nonumber.
\end{align}

As a practical matter, $P_o$ is large since the radar experiences an ideal power loss factor of $1/R^4$ \footnote{The power loss is greater considering other real world effects such as antenna losses, radar cross section etc.}. This deems the power transmitted to be large, in the orders of several MW to detect targets at ranges of  a few kilometers. With this in effect, it was shown in\cite{Setlurrangareport2014,SetlurTSP2016} that $\lambda=0$ when $P_o>\kappa$. When it is desired that the power transmitted be strictly equal to $P_o$, then the general solution presented in \eqref{rootsol} must be used.

{\bf Computational and hardware oriented compromises:} We note that the joint receiver and waveform design AM involves two matrix inversions at each iteration and is therefore computationally complex. Computational complexity may be reduced by performing projected gradient alternating descent on the problem, the details of which may be found in \cite{Setlurradar2015}. The objective in the projected gradient descent algorithm is to circumvent matrix inversion at each stage. A  tradeoff between achieving a lower objective and the number of iterations was observed in \cite{Setlurradar2015}.

As another compromise, $\lambda=0$ may {\it always} be chosen by making the power constraint redundant (inactive in Lagrange theory parlance), and then solving the design problem \cite{Setlurrangareport2014}. Subsequently, one would scale the solution to satisfy the power constraint with equality. This approach offers a small computational advantage but with some trade-off, as discussed subsequently. More importantly, however, this is appealing from a hardware standpoint since solving the equation in \eqref{rootsol} involves root finding of a nonlinear equation and may not be easily implementable in hardware.  In cases when an exact solution needs to be found -- or for example, when  in the radar case the matrix $\sum\limits_{q=1}^Q \mathbf{Z_q}( \mathbf{w_k})$ in the waveform design stage is rank deficient --  \eqref{rootsol} may then be used. 

{\bf Scaling the $\lambda=0$ solution:} If $(\mathbf{w}^{\ast},\mathbf{s}^{\ast})$ is a solution of the alternating minimization with $\lambda=0$ ( i.e. it is either a limit point or a solution which satisfies the desired objective tolerance). Then consider the following scaled solution, $(\frac{|| \mathbf{s}^{\ast}|| }{\sqrt{P_o}}\mathbf{w}^{\ast},\frac{\sqrt{P_o}}{|| \mathbf{s}^{\ast}||} \mathbf{s}^{\ast} )$. The scaled solution satisfies the power constraint as well as the capon constraint. Now since we can write $\mathbf{R}_{\bf c}(\mathbf{s})=\sum \limits _{q} \mathbf{A}_{\bf q} \mathbf{s} \mathbf{s}^H \mathbf{A}^H_{\bf q}$, the scaled solution has an identical clutter objective as the original solution. That is mathematically,
\begin{equation*}
\begin{aligned}
\frac{|| \mathbf{\mathbf{s}^{\ast}}||\mathbf{w}^{\ast H}}{\sqrt{P_o}} \mathbf{R}_{\bf c}\left( \frac{\sqrt{P_o}\mathbf{s^{\ast}}}{|| \mathbf{s}^{\ast}||} \right) \frac{|| \mathbf{\mathbf{s}^{\ast}}||\mathbf{w}^{\ast}}{\sqrt{P_o}}=\mathbf{w}^{\ast H} \mathbf{R}_{\bf c}(\mathbf{s^{\ast}})\mathbf{w}^{\ast}.
\end{aligned}
\end{equation*}
 However, the scaling does increase the noise plus interference response by  a factor of $|| \mathbf{s}^{\ast}||^2/P_o$ when compared to the original solution. In addition there is no guarantee that the scaled solution  satisfies the KKT's of the original problem. Nonetheless, if mitigating the clutter response or having it at a desirable level is appealing, then scaling is preferred.

\subsection{QCQP Formulation} We demonstrate a relation of the waveform design in \eqref{weightcomp2} to a quadratically constrained quadratic programming (QCQP) problem. At the $k$-th iteration, the waveform design is
\begin{align}\label{forqcqp1}
\min\limits_{\mathbf{s}} \mbox{ } &\mathbf{s}^H\mathbf{F}(\mathbf{w_{k}})\mathbf{s} \nonumber \\
\mbox{s. t . } \mbox{ } &\mathbf{s}^H\mathbf{y_{w}}_k=\kappa \\
 &|| \mathbf{s}||^2\leq P_o \nonumber
\end{align}
where we have $\mathbf{y_{w}}_k=\mathbf{G}^H\mathbf{w_k}$, and we intentionally write $\mathbf{F}$ as $\mathbf{F}(\mathbf{w_k})$ to reinforce the dependence on $\mathbf{w_k}$.
The weight vector must span some of the noise or interference and clutter subspaces, and hence $\mathbf{y_w}=\mathbf{G}^H\mathbf{w}$ must span at least some of these subspaces. Assume then, without loss of any generality, that it spans a $K$ dimensional subspace $\mathtt{U}$. We then decompose $\mathbb{C}^{N}$ as
 \begin{align} \label{subsp1}
 \mathbb{C}^{N}=\mathtt{U}\oplus \mathtt{V}, \quad \dim\{ \mathtt{U}\}=K, \quad \dim\{ \mathtt{V}\}=N-K.
 \end{align}
 In light of \eqref{subsp1} and the preceding facts,
 the vector $\mathbf{s}$ can be rewritten as
 \begin{align} \label{subsp2}
 \mathbf{s}=\mathbf{s}_\mathtt{U}+\mathbf{s}_\mathtt{V}
 =\mathbf{s}_{\mathbf{y_w}}+\mathbf{s}_{\mathbf{y_w}}^{\mathtt{U}}+\mathbf{s}_\mathtt{V}
 \end{align}
 where
 \begin{equation}\label{subsp3}
 \begin{aligned}
 &\mathbf{s}_\mathtt{U}\in\mbox{Span}\{ \mathtt{U}\}, \ \mathbf{s}_\mathtt{V}\in\mbox{Span}\{ \mathtt{V}\}, \
 \mathbf{s}_{\mathbf{y_w}}=\frac{<\mathbf{s}, \ \mathbf{y_w}>}{||\mathbf{y_w}||^2} \mathbf{y_w}\\
 &\mathbf{s}_{\bf y_w}^{\mathtt{U}}\in \mbox{Span}\{ \mathtt{U}\} \mbox{ but } \perp \, \mathbf{y_w}.
 \end{aligned}
 \end{equation}
 Using projection matrices, it is readily shown that
\begin{equation}\label{projmat1}
\begin{aligned}
&\mathbf{s}_{\mathtt{U}}=\mathbf{P_u}\mathbf{s},\ \mathbf{s}_{\mathtt{V}}=(\mathbf{I}-\mathbf{P_u})\mathbf{s}, \\
 &\mathbf{s}_{\bf y_w}=\mathbf{P_w}\mathbf{s}, \
 \mathbf{s}_{\bf y_w}^{\mathtt{U}}=(\mathbf{I}-\mathbf{P_w})\mathbf{P_u}\mathbf{s}
 \end{aligned}
\end{equation}
If $\mathtt{U}$ is made up of $\mathbf{U}=[\mathbf{u_1},\mathbf{u_2},\ldots,\mathbf{u_K}]\in\mathbb{C}^{N\times K}$ linearly independent vectors, then $\mathbf{P_u}=\mathbf{U} (\mathbf{U}^H\mathbf{U})^{-1}\mathbf{U}^H$ and $\mathbf{P_w}=\frac{\mathbf{y_w}\mathbf{y_w}^H}{|| \mathbf{y_w}||^2}$. 
Now using these arguments, the problem can be reformulated as
\begin{equation} \label{correctform}
\begin{aligned}
\min\limits_{\mathbf{s}_{y_w},\mathbf{s}_{y_w}^{\mathtt{U}},\mathbf{s}_{\mathbf{V}}}\,\,\, &\mathbf{s}^H \mathbf{F}(\mathbf{w_k})\mathbf{s}  \\
\mbox{s. t. } \,\,\, &\mathbf{s}^H \mathbf{y_w}=\kappa \\
& ||\mathbf{s}||^2 \leq P_o\\
&  \mathbf{s}
 =\mathbf{s}_{\mathbf{y_w}}+\mathbf{s}_{\mathbf{y_w}}^{\mathtt{U}}+\mathbf{s}_\mathtt{V}.
\end{aligned} 
\end{equation}

Consider a change of variables, $\mathbf{s}_\nu=\frac{\kappa}{<\mathbf{s}, \mathbf{y}_1>}\mathbf{s}_{\mathbf{y_w}}$, and $\mathbf{s}^{\mathtt{U}} _{\mathtt{V}}=\mathbf{s}_{\mathbf{y_w}}^{\mathtt{U}}+\mathbf{s}_\mathtt{V}$.  Further, since $\mathbb{C}^{N}=\mathbf{w}^\perp \oplus \mathbf{w}$, with $\mathbf{s}_{\mathtt{V}}^{\mathtt{U}} \in \mbox{Span} \{\ll \mathbf{w}^\perp \gg\}$, we readily have $\mathbf{P_w}\mathbf{P_u}=\mathbf{P_u}\mathbf{P_w}=\mathbf{P_w}$, where we use the notation $\ll \mathbf{w}^{\perp} \gg$ as the subspace orthogonal to $\mathbf{w}$. This is because $\mathbf{w}\subset\mathtt{U},\ll\mathbf{w}^{\perp}\gg\subset \mathtt{V}$. Hence the problem in \eqref{correctform} can be re-written as
\begin{equation} \label{correctform1}
\begin{aligned}
\min\limits_{\mathbf{s}_{\nu},\mathbf{s}^{\mathtt{U}}_{\mathbf{V}}}\,\,\, &\frac{|<\mathbf{s},\mathbf{y_w}>|^2}{|| \mathbf{y_w}||^4}\mathbf{y_w}^H \mathbf{F}(\mathbf{w_k})\mathbf{y_w} + \mathbf{s}^{\mathtt{U}H}_{\mathbf{V}} \mathbf{F}(\mathbf{w_k}) \mathbf{s}^{\mathtt{U}}_{\mathbf{V}}    \\
+& 2\mbox{Re}\{ \mathbf{s}^{\mathtt{U } H}_{\mathbf{V}} \mathbf{F}(\mathbf{w_k}) \frac{<\mathbf{s},\mathbf{y_w}>}{|| \mathbf{y_w}||^2} \mathbf{y_w}\} \\
\mbox{s. t. } \,\,\, &<\mathbf{s},\mathbf{y_w}>=\kappa \\
& ||\mathbf{s}^{\mathtt{U}}_{\mathtt{V}} ||^2 \leq P_o-\kappa^2/||\mathbf{y_1}||^2\\
&\mathbf{s}^{\mathtt{U}H}_{\mathtt{V}}\mathbf{y_w}=0 \\
&  \mathbf{s}
 =\frac{< \mathbf{s},\mathbf{y_1}>}{\kappa}\mathbf{s}_{\nu}+\mathbf{s}^{\mathtt{U}}_{\mathtt{V}}.
\end{aligned} 
\end{equation}

If and only if a solution exists to \eqref{correctform1}, then this solution also is a solution to
\begin{equation} \label{correctform1a}
\begin{aligned}
\min\limits_{\mathbf{s}^{\mathtt{U}}_{\mathbf{V}}}\mbox{ } &\frac{\kappa^2}{|| \mathbf{y_w}||^4}\mathbf{y_w}^H \mathbf{F}(\mathbf{w_k})\mathbf{y_w} + \mathbf{s}^{\mathtt{U}H}_{\mathbf{V}} \mathbf{F}(\mathbf{w_k}) \mathbf{s}^{\mathtt{U}}_{\mathbf{V}}    \\
+& 2\frac{\kappa}{|| \mathbf{y_w}||^2}\mbox{Re}\{ \mathbf{s}^{\mathtt{U } H}_{\mathbf{V}} \mathbf{F}(\mathbf{w_k}) \mathbf{y_w}\} \\
\mbox{s. t. } \,\,\, & ||\mathbf{s}^{\mathtt{U}}_{\mathtt{V}} ||^2 \leq P_o-\kappa^2/||\mathbf{y_1}||^2\\
&\mathbf{s}^{\mathtt{U}H}_{\mathtt{V}}\mathbf{y_w}=0.
\end{aligned} 
\end{equation}

We can simplify \eqref{correctform1a} further, consider an arbitrary $\mathbf{q}\in\mathbb{C}^N$, the vector $\mathbf{s}^{\mathtt{U}}_{\mathbf{V}}=\mathbf{P}^{\perp}_{\bf w}\mathbf{q}=(\mathbf{I}-\mathbf{P_w})\mathbf{q}$ will also always satisfy the last constraint. Using this fact and ignoring the constant term of the objective in \eqref{correctform1a}, we have,
\begin{equation} \label{correctform2}
\begin{aligned}
\min\limits_{\mathbf{q}}\mbox{ } &\mathbf{q}^H \mathbf{P}^{\perp}_{\bf w}\mathbf{F}(\mathbf{w_k}) \mathbf{P}^{\perp}_{\bf w} \mathbf{q}+2\frac{\kappa}{|| \mathbf{y_w}||^2}\mbox{Re}\{ \mathbf{q}^H \mathbf{P}^{\perp}_{\bf w} \mathbf{F}(\mathbf{w_k}) \mathbf{y_w}\}  \\
\mbox{s. t. } \,\,\, 
& ||\mathbf{P}^{\perp}_{\bf w}\mathbf{q}||^2 \leq P_o-\frac{\kappa^2}{ ||\mathbf{y_w}||^2}
\end{aligned} 
\end{equation}
The solution to \eqref{correctform2} is readily shown to be,
\begin{equation} \label{correcform3}
\begin{aligned}
&\gamma \left(\frac{\kappa^2}{|| \mathbf{y_w}||^4}\mathbf{y_w}^H\mathbf{A}^H(\gamma) \mathbf{P}^{\perp}_{\bf w} \mathbf{A}(\gamma) \mathbf{y_w}
-P_o+\frac{\kappa^2}{||\mathbf{y_w}||^2} \right) =0 \\
&\mathbf{A}(\gamma)= (\mathbf{P}^{\perp}_{\bf w}\mathbf{F}(\mathbf{w_k}) \mathbf{P}^{\perp}_{\bf w}+\gamma\mathbf{P}^{\perp}_{\bf w})^{\dagger} \mathbf{P}^{\perp}_{\bf w}\mathbf{F}(\mathbf{w_k}), \mbox{ } \gamma \geq 0.
\end{aligned}
\end{equation}
where $\gamma$ is another Lagrange parameter and $(\cdot)^{\dagger}$ denotes the pseudoinverse of a matrix. The waveform design solution at the $k$-th iteration of the AM is given by
\begin{equation}  \label{finalwavwsol}
\begin{aligned}
\mathbf{s_k}&=\mathbf{P}^{\perp}_{\bf w} \mathbf{q}(\gamma^\ast)+\kappa\frac{\mathbf{y_w}}{|| \mathbf{y_w}||^2} \\
\mathbf{q}(\gamma)&=-\frac{\kappa}{|| \mathbf{y_w}||^2} \mathbf{A}(\gamma)\mathbf{y_w}
\end{aligned}
\end{equation}
where $\gamma^{\ast}$ is the solution of \eqref{correcform3}.

Similar arguments on $\gamma$ may be made as in $\lambda$ for the original problem. That is, $\gamma=0$ may be chosen to make computational and hardware oriented compromises.
\subsection{SDP Formulation}
A semi-definite program (SDP) formulation to \eqref{correctform2} is readily seen. Consider the dual problem of \eqref{correctform2}, given by
\begin{align*} 
\inf\limits_{\mathbf{q}} \ &\mathbf{q}^H ( \mathbf{P}^{\perp}_{\bf w}\mathbf{F}(\mathbf{w_k}) \mathbf{P}^{\perp}_{\bf w} + \alpha \mathbf{P}^{\perp}_{\bf w} )\mathbf{q}+2\frac{\kappa}{|| \mathbf{y_w}||^2}\mbox{Re}\{ \mathbf{q}^H \mathbf{P}^{\perp}_{\bf w} \mathbf{F}(\mathbf{w_k}) \mathbf{y_w}\}  \nonumber \\
& +\frac{\alpha\kappa}{||\mathbf{y_w} ||^2}-\alpha P_o 
\end{align*}
\begin{equation*}
=\begin{cases}
 \frac{\alpha \kappa}{|| \mathbf{y_w}||^2} -\alpha P_o-\frac{\kappa^2}{|| \mathbf{y_w}||^4} \mathbf{b}^H\mathbf{B}^{\dagger}(\alpha)\mathbf{b}^H  \ &  \begin{aligned} &\mathbf{B}(\alpha)\succeq 0, \\
 &\mathbf{b}\notin \mbox{Null}\left( \mathbf{B}(\alpha)\right) \end{aligned} \\
 & \\
 -\infty  \ & \mbox{otherwise}
\end{cases}
\end{equation*}
where $\mathbf{B}(\alpha) = \mathbf{P}^{\perp}_{\bf w}(\mathbf{F}(\mathbf{w_k}) + \alpha \mathbf{P}^{\perp}_{\bf w})\mathbf{P}^{\perp}_{\bf w},\;\;\mathbf{b} = \mathbf{P}^{\perp}_{\bf w} \mathbf{F}(\mathbf{w_k}) \mathbf{y_w}$.

Using the approach of Shor \cite{Shor1987}, we have the SDP dual formulation 
\begin{align} \label{sdpdual1}
\max \limits_{\alpha, \beta} &\ \beta \nonumber \\
\mbox{s. t . } &\begin{bmatrix} 
\mathbf{B}(\alpha) & \frac{\kappa}{|| \mathbf{y_w}||^2}\mathbf{b} \\
\frac{\kappa}{|| \mathbf{y_w}||^2}\mathbf{b}^H &  \frac{\alpha \kappa}{|| \mathbf{y_w}||^2} -\alpha P_o-\beta
\end{bmatrix} \succeq 0\\
&\alpha\geq 0 \nonumber.
\end{align}
Alternatively, after a change of variables,
\begin{align} \label{sdpdual2}
\max \limits_{\alpha, \beta} &\ \beta+\frac{\alpha \kappa}{|| \mathbf{y_w}||^2} -\alpha P_o  \nonumber \\
\mbox{s. t . } &\begin{bmatrix} 
\mathbf{B}(\alpha) & \frac{\kappa}{|| \mathbf{y_w}||^2}\mathbf{b} \\
\frac{\kappa}{|| \mathbf{y_w}||^2}\mathbf{b}^H &  -\beta
\end{bmatrix} \succeq 0 \\
&\alpha\geq 0 \nonumber.
\end{align}

{\bf Primal SDP relaxation}
 Another well known approach to SDP relaxation is by rewriting \eqref{correctform2} as
 \begin{equation}
\begin{aligned} \label{sdpprimal1}
\min\limits_{\mathbf{q}} \ &\mbox{Tr}  \left\lbrace 
\begin{bmatrix}
\mathbf{qq}^H & \mathbf{q} \\
\mathbf{q}^H &1
\end{bmatrix}
\begin{bmatrix}
\mathbf{B}(0) & \frac{\kappa \mathbf{b}}{|| \mathbf{y_w}||^2} \\
\frac{\kappa \mathbf{b}^H}{|| \mathbf{y_w}||^2} & 0
\end{bmatrix}
\right\rbrace \\
\mbox{s. t. } & \mbox{Tr}  \left\lbrace 
\begin{bmatrix}
\mathbf{qq}^H & \mathbf{q} \\
\mathbf{q}^H &1
\end{bmatrix}
\begin{bmatrix}
\mathbf{P}^{\perp}_{\bf w}  & 0 \\
0 & 0
\end{bmatrix}
\right\rbrace \leq P_o-\frac{\kappa^2}{ ||\mathbf{y_w}||^2}
\end{aligned}
\end{equation}

From \eqref{sdpprimal1}, we recognize immediately that if we substitute $\mathbf{Q}=\left[\begin{smallmatrix}
\mathbf{qq}^H & \mathbf{q} \\
\mathbf{q}^H &1
\end{smallmatrix} \right]$ and with $\mathbf{Q}\succeq0$, we have the SDP primal relaxation 
\begin{equation}
\begin{aligned} \label{sdpprimal2}
\min\limits_{\mathbf{Q}} \ &\mbox{Tr}  \left\lbrace 
\mathbf{Q}
\begin{bmatrix}
\mathbf{B}(0) & \frac{\kappa \mathbf{b}}{|| \mathbf{y_w}||^2} \\
\frac{\kappa \mathbf{b}^H}{|| \mathbf{y_w}||^2} & 0
\end{bmatrix}
\right\rbrace \\
\mbox{s. t. } & \mbox{Tr}  \left\lbrace 
\mathbf{Q}
\begin{bmatrix}
\mathbf{P}^{\perp}_{\bf w}  & 0 \\
0 & 0
\end{bmatrix}
\right\rbrace \leq P_o-\frac{\kappa^2}{ ||\mathbf{y_w}||^2} \\
&\mathbf{Q}\succeq 0, \mbox{ with } \mathbf{Q}(N,N)=1.
\end{aligned}
\end{equation}

\begin{remk}[{\it Strong Duality}] \label{remark1}
It can be readily shown that, given a sign change for the variable $\beta$, \eqref{sdpdual2} is \emph{also} the dual problem for \eqref{sdpprimal2}. Thus, we can claim the following (see, e.g., \cite[Appendix B]{Boyd2004} for the real variable case): Since the primal SDP \eqref{sdpprimal2} is a relaxation of the QCQP \eqref{correctform2}, then its optimal value is a lower bound on the QCQP's optimal value; that is, $\nu^{o}_{QCQP} \geq \nu^{o}_{SDP-P}$. Furthermore, as a dual problem for \eqref{sdpprimal2} , the dual SDP \eqref{sdpdual2}'s value is a lower bound on the primal SDP, which thus implies $\nu^{o}_{QCQP} \geq \nu^{o}_{SDP-P} \geq \nu^{o}_{SDP-D}$. Now, if there is a strictly feasible solution for \eqref{correctform2}, then strong duality between \eqref{correctform2} and \eqref{sdpdual2} holds, and $\nu^{o}_{SDP-D} = \nu^{o}_{QCQP}$. Clearly then, $\nu^{o}_{SDP-P} = \nu^{o}_{QCQP}$ as well, which implies strong duality holds between \eqref{correctform2}, \eqref{sdpdual2}, and \eqref{sdpprimal2}. In fact, any rank-1 optimal solution of \eqref{sdpprimal2} is therefore optimal for \eqref{correctform2}.
\end{remk}
\subsection{Least Squares on Hyperellipses}
We can show that \eqref{correctform2} can be rewritten as a constrained least squares, specifically, least squares constrained to lie on  a hyperellipsoid. This can be seen since $\mathbf{F}(\mathbf{w_k})$ is Hermitian and permits a square root factorization, $\mathbf{B}(0)=\sqrt{\mathbf{F}(\mathbf{w_k})}^H \sqrt{\mathbf{F}(\mathbf{w_k})}$. Let $\mathbf{C}=\mathbf{P}^{\perp}_{\bf w}\sqrt{\mathbf{F}(\mathbf{w_k})}^H$ and $\mathbf{d}=-\frac{\kappa}{|| \mathbf{y_w}||^2}\sqrt{\mathbf{F}(\mathbf{w_k})}\mathbf{y_w}$. Then, \eqref{correctform2} can be written as,
\begin{equation} \label{lsconstr}
\begin{aligned}
\min\limits_{\mathbf{q}}\mbox{ } & ||\mathbf{Cq}-\mathbf{d} ||^2 \\
\mbox{s. t. } \,\,\, 
& ||\mathbf{P}^{\perp}_{\bf w}\mathbf{q}||^2 \leq P_o-\frac{\kappa^2}{ ||\mathbf{y_w}||^2}
\end{aligned} .
\end{equation}
This is a constrained least squares problem in standard form, with the constraint being a hyperellipse. A solution to \eqref{lsconstr} is straightforward but is not presented here due to space constraints. The SVD can be used very efficiently to solve \eqref{lsconstr}, see for example \cite{Golub1996}. 

We note now that the optimal solution $\mathbf{q}^{\ast}$ from any one of \eqref{sdpdual1}, \eqref{sdpdual2} and the one derived from \eqref{sdpprimal2} may be substituted in \eqref{finalwavwsol} to obtain the optimal $\mathbf{s}_{k}^{\ast}$.
\section{Convergence of AM for Bi-convex Optimization}
In this section, we discuss the conditions and assumptions under which the AM algorithm converges for the bi-convex (see Definition~\ref{def1}) optimization in \eqref{biconvex1}. Using these assumptions and conditions, we prove that iff and only if limit points exist for a bi-convex problem, then those limit points are also stationary. We delineate the conditions  necessary for convergence next.

{\textbf{Condition \textit{C\textsubscript{1}:}}} We first assume that a functional relationship exists between the iterates at the $k$-th iteration and their previous counterparts. That is, assume
\begin{equation}\label{functional1}
\begin{aligned}
\mathbf{x_k}=f_1(\mathbf{y_{k-1}})=f_1(f_2(\mathbf{x_{k-1}}))
:=f_y(\mathbf{x_{k-1}})\\
\mathbf{y_k}=f_2(\mathbf{x_{k-1}}):=f_x(\mathbf{x_{k-1}}), k=1,2,\ldots
\end{aligned}
\end{equation}
where $f_1:\mathbb{F}^M\rightarrow \mathbb{F}^{N}$ and $f_2:\mathbb{F}^{N}\rightarrow\mathbb{F}^M$, and $\mathbf{x_0}\in\mathcal{X}_0,\,\mathbf{y_0}\in\mathcal{Y}_0$. 

Most well defined and practical bi-convex problems may have such functional relationships. For our radar problem, it is evident from \eqref{weightcomp} that such functional relationships definitely exist. 

Define the sequence of iterates from the AM algorithm as $\mathbf{x_k},\mathbf{y_k},k=1,2,\ldots$.

{\textbf{Condition \textit{C\textsubscript{2}:}} } The constraint sets are an explicit function of the previous iterates: $\mathcal{A}_k(\mathbf{\mathbf{x_{k-1}})}:=\left\lbrace\mathbf{x} | g_i(\mathbf{x},f_x(\mathbf{x_{k-1}} ))\leq 0,\, h_j( \mathbf{x},f_x(\mathbf{x_{k-1}} ))=0\right\rbrace$,  and  $\mathcal{B}_k(\mathbf{\mathbf{x_{k-1}})}:=\left\lbrace\mathbf{y} | g_i(\mathbf{x_{k-1}},\mathbf{y} )\leq 0, h_j( \mathbf{x_{k-1}},\mathbf{y} )=0\right\rbrace$, with $i=1,2,\ldots, j=1,2,\ldots$. 

From here onward,  for simplicity of notation, we will drop the explicit dependence of the previous iterates on these sets, and denote them simply as $\mathcal{A}_k,\mathcal{B}_k$. Therefore, we can rewrite \eqref{ambiconvex} succinctly as
\begin{align}\label{ambiconvex3}
\mathbf{x_k}=\argmin \limits_{\mathbf{x}\in\mathcal{A}_k} f(\mathbf{x},\mathbf{y_{k-1}}), \ \mathbf{y_k}=\argmin \limits_{\mathbf{y}\in\mathcal{B}_k} f(\mathbf{x_{k-1}},\mathbf{y}).
\end{align}
Now assume that the sequences $\{ \mathbf{x_k}\}$ \& $\{ \mathbf{y_k}\}$ and each of their corresponding subsequences have a finite limit point. Further assume that $f$ is uniformly continuous everywhere, and that its domain includes the Cartesian product of two metrizable supersets.

The definition of Hausdorff distance and the diameter of a set will also prove to be useful.
\begin{defn}[Hausdorff distance] \label{def2}
Let $\mathcal{A}, \mathcal{B}$ be two sets in some metric space $(\mathcal{X}.d)$, the Hausdorff distance is defined as,
\begin{align}
d_{\mathcal{H} }(\mathcal{A},\mathcal{B}) :=\sup \limits_{\mathbf{a}\in \mathcal{A}} \ \inf\limits_{\mathbf{b} \in \mathcal{B}} d(\mathbf{a},\mathbf{b}) \vee \sup \limits_{\mathbf{b}\in \mathcal{B}} \ \inf\limits_{\mathbf{a} \in \mathcal{A}} d(\mathbf{a},\mathbf{b}).
\end{align}
Where as usual, $d(\mathbf{a},\mathbf{b})=|| \mathbf{a}-\mathbf{b}||$. One can think of the Hausdorff distance as metric which measures similarity of two sets, $\mathcal{A},\mathcal{B}$.
\end{defn}
\begin{defn}[Diameter of  a set] \label{def3}
A set $\mathcal{A}$ has a diameter, $D_\mathcal{A}=\sup \ d(\mathbf{a},\mathbf{b})$, for $\mathbf{a}\in\mathcal{A},\mathbf{b} \in\mathcal{A}$.
\end{defn}

With respect to notation, we denote the convex hull of a set $\mathcal{A}$ as $\mbox{Conv}(\mathcal{A})$, and the closure of a set as $\mbox{Cl}(\mathcal{A})$. Now define the sets, with a slight abuse of notation,
\begin{equation*}
\begin{aligned}
\mathcal{C}_{\mathcal{X}_0} :=\bigcup \limits_{\mathbf{x_0} \in\mathcal{X}_0}\mathcal{C}_{\mathbf{x_0}}, \mbox{   }
\mathcal{D}_{\mathcal{Y}_0} :=\bigcup \limits_{\mathbf{y_0}\in\mathcal{Y}_0 } \mathcal{D}_{\mathbf{y_0}}
\end{aligned}
\end{equation*}
where 
\begin{equation}\label{set1}
\begin{aligned}
\mathcal{C}_{\mathbf{x_0}} &=\{\mathbf{x_0}\} \; \cup \; \left\{ \bigcup \limits_{k\in\mathbb{Z}^{+}}\{\mathbf{x_k} | \mathbf{x_k}= \argmin \limits_{\mathbf{x}\in\mathcal{A}_k} f(\mathbf{x},\mathbf{y_{k-1}})\}\right\} \\
\mathcal{D}_{\mathbf{y_0}} &=\{\mathbf{y_0}\} \; \cup \; \left\{\bigcup \limits_{k\in\mathbb{Z}^{+}}\{\mathbf{y_k}  | \mathbf{y_k}=\argmin \limits_{\mathbf{y}\in\mathcal{B}_k} f(\mathbf{x_{k-1}},\mathbf{y}) \}\right\}
\end{aligned}
\end{equation}
and where $\mathbb{Z}^{+}=\{1,2,3\ldots\}$. The sets $\mathcal{C}_{\mathcal{X}_0},\mathcal{D}_{\mathcal{Y}_0}$ are the set of {\it all} $\mathbf{x}, \mathbf{y}$ iterates for every $\mathbf{x}_0\in\mathcal{X}_0$, $\mathbf{y}_0\in\mathcal{Y}_0$, respectively. 

Define the two convex closures $\mathring{\mathcal{C}}_{\mathcal{X}_0}, \mathring{\mathcal{D}}_{\mathcal{Y}_0}$ as
\begin{align}\label{defconsets}
\mathring{\mathcal{C}}_{\mathcal{X}_0}=\mbox{Cl}(\mbox{Conv}(\mathcal{C}_{\mathcal{X}_0})), \; \mathring{\mathcal{D}}_{\mathcal{Y}_0}=\mbox{Cl}(\mbox{Conv}(\mathcal{D}_{\mathcal{Y}_0}))
\end{align}
and endow the topology $\mathscr{T}_x$ to $\mathring{\mathcal{C}}_{\mathcal{X}_0}$, and the topology $\mathscr{T}_y$ to $\mathring{\mathcal{D}}_{\mathcal{Y}_0}$, both of which are induced by the metric $d(\cdot,\cdot)$. We are not interested in arbitrary sequences of some arbitrary subsets $\bar{\mathcal{C}}_x\subset\mathring{\mathcal{C}}_{\mathcal{X}_0}$ and $\bar{\mathcal{D}}_y\subset\mathring{\mathcal{D}}_{\mathcal{X}_0}$ but only those sequences, $\{ \mathbf{x}^{c}_{\bf k}\}$ and $\{\mathbf{y}^{d}_{\bf k} \}$ such that,
\begin{align}\label{sequence1}
\mathbf{x}^{c}_{\bf k}=\argmin \limits_{\mathbf{x} \in\mathring{\mathcal{C}}_{\mathcal{X}_0}} f(\mathbf{x},\mathbf{y}^{d}_{\bf k-1}), \ \mathbf{y}^{d}_{\bf k}=\argmin \limits_{\mathbf{y} \in \mathring{\mathcal{D}}_{\mathcal{Y}_0}} f(\mathbf{x}^{c}_{\bf k-1},\mathbf{y}).
\end{align}
{ \it Assumption 1 :} Assume that the sequences $\{ \mathbf{x}^{c}_{\bf k}\}$ and $\{\mathbf{y}^{d}_{\bf k} \}$  have a limit point identical to the sequences, $\{\mathbf{x_k}\}$ and $\{ \mathbf{y_k} \}$, denoted as $\mathbf{x}^{\ast}$ and $\mathbf{y}^{\ast}$, respectively, and for any initialization $\mathbf{x}_{\bf 0} \in\mathcal{X}_0$, $\mathbf{y}_{\bf 0} \in\mathcal{Y}_0$.

This assumption is motivated by considering $D_ {\mbox{Conv}(\mathcal{C}_{\mathcal{X}_0})} \leq \epsilon_1$, and $D_{\mbox{Conv}(\mathcal{D}_{\mathcal{Y}_0})  }\leq \epsilon_2$. Typically, both $\epsilon_1$ and $\epsilon_2$ are small but positive. 

We now prove that iff a limit point $(\mathbf{x}^{\ast},\mathbf{y}^{\ast} )$ exists, then this limit point is also stationary. A sketch of the proof is provided next.
\begin{thm}[Convergence of AM] \label{them1}Let Assumption 1 be satisfied and if
\begin{enumerate}
\item  functional relationships $\mathbf{x}_k=f_y(\mathbf{x}_{k-1})$ and $\mathbf{y}_k=f_x(\mathbf{x}_{k-1})$ exist and the minimizers at each stage of the AM are unique and,
\item for $k\rightarrow \infty$, the constraint sets are stationary, i.e $\lim \limits_{k \rightarrow\infty} d_{\mathcal{H}}(\mathcal{A}_k,\mathcal{A})\rightarrow 0$ and $\lim \limits_{k \rightarrow\infty} d_{\mathcal{H}}(\mathcal{B}_k,\mathcal{B})\rightarrow 0$,
\end{enumerate}
then if a limit point $(\mathbf{x}^{\ast},\mathbf{y}^{\ast} )$ exists, then it is also a stationary point.
\end{thm}
\begin{proof}
Since we require Assumption 1 to be satisfied, we will consider only those special limit points which satisfy Assumption 1. 

The functional relationships $f_x, f_y$ between the iterates are necessary for two reasons. First, they ensure simultaneous convergence in both the $\mathbf{x}$ and $\mathbf{y}$ iterates. If such a functional relationship did not exist, convergence in one would not influence convergence in the other, which might lead to cycling \& a failure to converge entirely. For example, consider the $k_o$th iteration of an algorithm without such a dependence. Assume {\it hypothetically} that the sequence beginning with the iterate $\mathbf{y}_{k_o}$ converges to the limit point $\mathbf{y}^{\ast}$, i.e. $\mathbf{y}_{k_o+n}\rightarrow  \mathbf{y}^{\ast}$ as $n\rightarrow \infty$. If the functional relationship did not exist, then the next $\mathbf{x}$ iterate, $\mathbf{x}_{k_o+1}$, is not guaranteed to be in a convergent sequence to the limit point $\mathbf{x}^{\ast}$. Furthermore then, the {\it hypothesis} is proven false, and there will almost certainly then exist a $k > k_o$ where a sequence starting with $\mathbf{y}_k$ will \emph{not} converge to $\mathbf{y}^{\ast}$. 

Second, these functional relationships  along with condition \textit{C\textsubscript{2}} enforce the dependency of the constraint sets $\mathcal{A}_k$ and $\mathcal{B}_k$ on the previous iterates, which means these sets are not arbitrary. Additionally then, convergence in the iterates will also enforce the stationarity of the constraint sets $\lim\limits_{k\rightarrow \infty} \mathcal{A}_k\rightarrow\mathcal{A} $, and $\lim\limits_{k\rightarrow \infty} \mathcal{B}_k\rightarrow \mathcal{B}$. If these sets do not converge, then there is no guarantee that the iterates themselves converge. 

Now consider the following optimization problem: 
\begin{align} \label{minfinal} 
\min \limits_{ \mathbf{x} \in \mathring{\mathcal{C}}_{\mathcal{X}_0}, \mathbf{y}\in \mathring{\mathcal{D}}_{\mathcal{Y}_0} } f(\mathbf{x},\mathbf{y})
\end{align}

The solution to \eqref{minfinal} is also $(\mathbf{x}^{\ast}, \mathbf{y}^{\ast})$. Assume that this is not true, and there is another limit point, $(\mathbf{x}^{\ast \blacklozenge},\mathbf{y}^{\ast \blacklozenge})=\inf \limits_{ \mathbf{x} \in \mathring{\mathcal{C}}_{\mathcal{X}_0}, \mathbf{y}\in \mathring{\mathcal{D}}_{\mathcal{Y}_0} } f(\mathbf{x},\mathbf{y})$. then since it belongs to the convex hull, $\exists\, \alpha_k,k=1,2,\ldots, \;\sum \limits_{i}\alpha_k=1$ and $\exists \beta_i,i=1,2,\ldots,\; \sum \limits_{i}\beta_k=1$, such that
\begin{align}
&f(\mathbf{x}^{\ast \blacklozenge},\mathbf{y}^{\ast \blacklozenge})=f\left( \sum \limits_{k} \alpha_k  \mathbf{x_k},\ \sum \limits_{k} \beta_k  \mathbf{y_k}\right) \nonumber \\
&\leq \sum \limits_{k} \alpha_k f\left(  \mathbf{x_k},\ \sum \limits_{k} \beta_k  \mathbf{y_k}\right)\\
&\leq \alpha \  \inf \limits_{\mathbf{x_k} }f\left( \mathbf{x_k}, \ \sum \limits_{k} \beta_k  \mathbf{y_k}\right) + (1-\alpha) \ \sup  \limits_{\mathbf{x_k} }f\left( \mathbf{x_k}, \ \sum \limits_{k} \beta_k  \mathbf{y_k}\right) \nonumber \\
&\implies f(\mathbf{x}^{\ast \blacklozenge},\mathbf{y}^{\ast \blacklozenge}) \leq \inf \limits_{\mathbf{x_k} }f\left( \mathbf{x_k}, \ \sum \limits_{k} \beta_k  \mathbf{y_k}\right) \nonumber
\end{align}
where, in the first inequality, we have used the convexity of $f(\mathbf{x},\mathbf{y})$ for a fixed $\mathbf{y}$, and the second inequality follows from the Carath{\'e }odory's theorem on the real axis. The last inequality follows as a special case for $\alpha=1$. Now starting from the last inequality and following the same procedure as before, we have 
\begin{align*}
f(\mathbf{x}^{\ast \blacklozenge},\mathbf{y}^{\ast \blacklozenge})&\leq\inf \limits_{\mathbf{x_k}, \mathbf{y_k} }f( \mathbf{x_k}, \mathbf{y_k}) \\
&=f( \mathbf{x}^{\ast},\mathbf{y}^{\ast}).
\end{align*}

This is a contradiction since, by Assumption 1, these sequences must have identical limit points. Therefore, $ \mathbf{x}^{\ast \blacklozenge}=\mathbf{x}^{\ast }$ and $\mathbf{y}^{\ast \blacklozenge} =\mathbf{y}^{\ast} $
We can now apply the same technique as in \cite[Prop. 2.7.1]{Bertsekas1999} 
to the problem in \eqref{minfinal} by using the AM algorithm on this modified optimization problem and demonstrate that limit point is also a stationary point.
\end{proof}

\section{Simulations}

In this section, we validate the concurrence of AM, the SDP and QCQP formulation given in Section 3 through a simulated example. Additionally, we will show that any of these algorithms, if appropriately rescaled to satisfy a power equality constraint, both outperforms a version of the technique in \cite{Demaio2013} and asymptotically approaches the optimal value of the relaxed problem given in \cite{ORourkeAFRLTR2016}.

In both of these cases, we will let $\kappa = P_o = 1$ for two reasons: first, to demonstrate the inherent tradeoff in restricting $\lambda = 0$ when the aforementioned condition is not satisfied, and second, to facilitate comparison with the algorithm in \cite{Demaio2013} where these particular values are implicitly assumed. 

The scenario presented to the solvers is as follows: We assume the radar transmits $L = 8$ pulses of $N = 8$ samples each, and receieves them with a uniform linear array of $M = 5$ elements on a moving platform. We assume that the signal independent noise covariance matrix $\mathbf{R_{n}}$ is a Toeplitz matrix with the $(i,j)$th element equal to $\exp(-0.005|i-j|)$. A single interferer is present at the azimuth elevation pair (0.3941, $\pi/3$) radians with a fast time-slow time correlation given by $\exp(0.02n)$. The target is located at the azimuth-elevation pair (0, $\pi/3$) radians, moving at a relative normalized Doppler of -0.1443 to the platform.  The clutter is described by 25 statistically independent patches whose nominal phase centers are at azimuths linearly spaced in the interval $[-\pi/2, \pi/2]$ radians and an elevation angle of 0.3 radians. Additional parameters are as in \cite{SetlurTSP2016}.

First, we compare the alternating minimization (AM) with the theoretically equivalent QCQP in \eqref{correctform1} and projected SDP in \eqref{sdpprimal1}. 
All algorithms were terminated after 20 iterations and the parameter $\lambda$ was directly determined by a line search at each iteration. 
Results of an example convergence run are demonstrated in Figure 1, where we compare them with the optimal value of the joint relaxed biquadratic progam (RBQP) proposed in \cite{ORourkeAFRLTR2016}. 
Clearly, when fully implemented, AM, the QCQP method, and Projected SDP (the first set of curves) coincide exactly and converge rather quickly to a certain limit point in a manner reminiscent of Theorem 1. 
Additionally, there is an expected gap between the optimal values of the original problem and its relaxation.
The second set of curves shows the equivalent cost if the power constraint was always satisfied with equality, which we enforce by rescaling the optimal $\mathbf{w}, \mathbf{s}$ pairs at each iteration such that $||\mathbf{s}_{k}||^2 = P_o$ and the Capon constraint is maintained.
This is important because the iterative solvers we propose prefer to sequentially lower the power until convergence, which makes comparison with techniques that enforce power limitations with equality difficult. 
As noted above, however, these rescaled iterates may not be KKT points of the originally considered problem. 
Unsurprisingly, the three algorithms again coincide; however, unlike the inequality constrained iterates, the rescaled iterates clearly converge to an asymptote given by the optimal value of the relaxed problem, despite not being guaranteed to satisfy the original problem's KKTs.
Hence, we have verified the equivalence and convergence properties of the aforementioned problems.
\begin{figure}[!ht]
	\centering
	\includegraphics[width=\columnwidth]{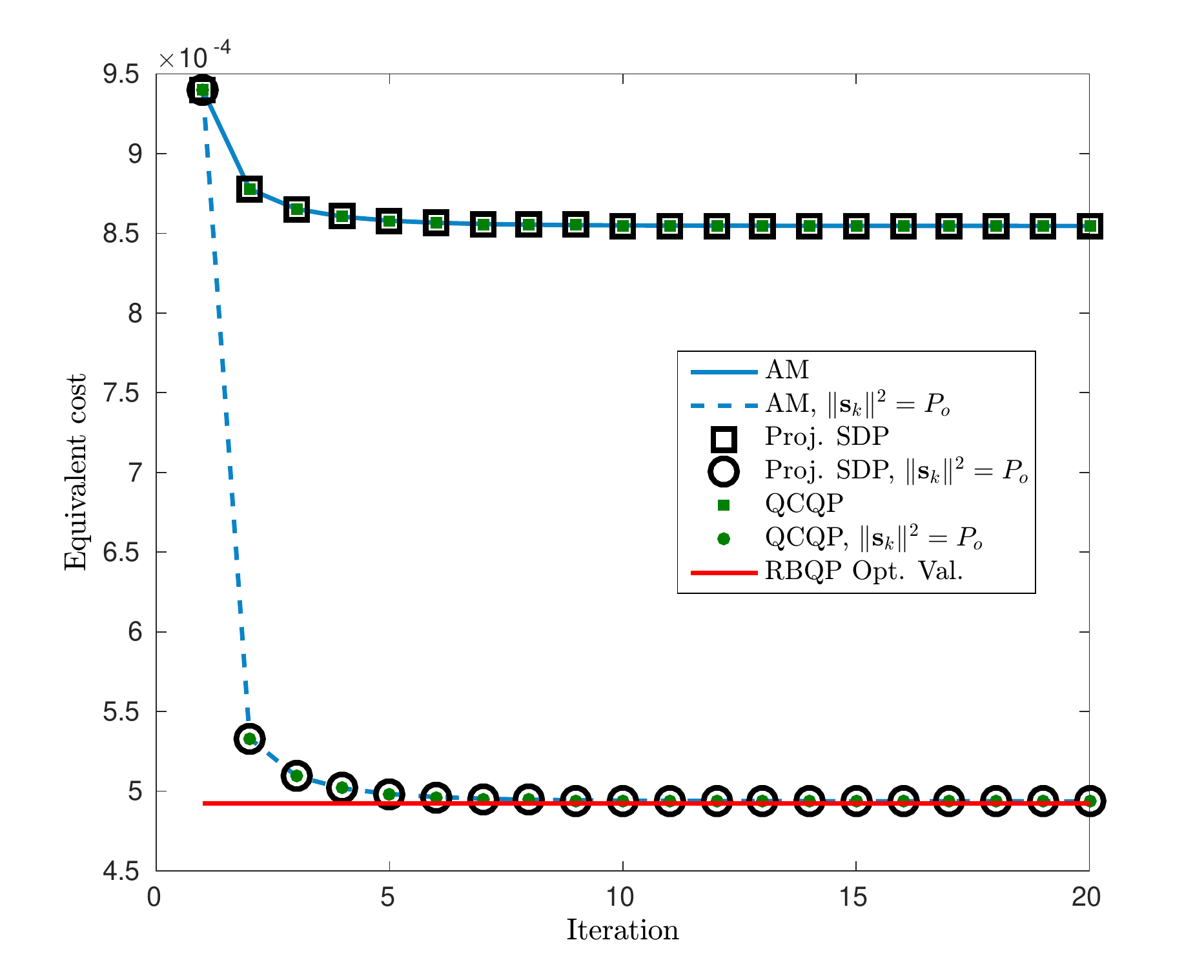}
	\caption{Convergence comparison of AM, QCQP, and projected SDP (unscaled and scaled).}
\end{figure}

Next, we compare the convergence properties of the equivalent methods above with two other algorithms. The first algorithm is a suboptimal form of projected SDP where the Lagrange multiplier $\lambda$ is set to zero at each step. 
This solution is suboptimal since $\lambda = 0$ does not necessarily satisfy the KKTs when $P_o = \kappa$.
Like the optimal version above, this version is terminated after 20 iterations.
The second algorithm is a modified version of the algorithm in \cite{Demaio2013} (henceforth, AA2) where the similarity constraint is removed. 
While AA2 nominally maximizes SINR, we can make a reasonable comparison of the equivalent objective for the other problems by examining the reciprocal. 
In this case, we set the convergence parameter $\epsilon = 1$ -- as convergence tolerances go, this may seem rather large, but it is our experience that even this can lead to extended runtimes for the algorithm.
For comparison, we examine the mean objective value $\hat{\nu}$ attained by each algorithm (rescaled, if necessary) after 20 iterations or, in the case of AA2, when the algorithm was deemed to have converged if it occurred before 20 iterations. 
This was achieved by averaging the results of 50 Monte Carlo trials, with each algorithm initialized with a given randomly-generated signal for each trial.
The results are presented in Table 1, where we express the resultant objective value in multiples of the optimal value of RBQP $\nu_{RBQP}^{\star}$ (here, $4.21 \times 10^{-4}$), which we have previously established as an empirical convergence asymptote.
\begin{table}[]
	\centering
	\caption{Mean convergence comparison of algorithms.}
	\label{tab:mc_convergence}
	\begin{tabular}{|c|c|}
		\hline
		Algorithm            & $\hat{\nu}/\nu_{RBQP}^{\star}$ \\ \hline
		\hline
		AM/QCQP              & 2.12  \\ \hline
		SDP                  & 2.06  \\ \hline
		SDP, $\lambda = 0$   & 1.33  \\ \hline
		AM/QCQP/SDP Rescaled & 1.01  \\ \hline
		AA2                  & 2.16  \\ \hline
	\end{tabular}
\end{table}

It is clear that AA2 to an effectively higher limit point than any of the generally equivalent forms above or their more comparable rescaled counterparts. This is also true for the suboptimal projected SDP; however, even this method outperforms AA2 in our experiments. 
Furthermore, suboptimal projected SDP requires minimal rescaling (that is, $||\mathbf{s}_{k}||^{2} \approx P_o$ for all iterations). Hence, it may be a more viable option than suboptimal AM for times when $P_o \approx \kappa$, which frequently has iterates exceed the power constraint if $\lambda$ is forced to zero and, when rescaled, can produce solutions with higher costs than any of the methods listed herein. 




\bibliographystyle{IEEEtran}
\bibliography{refs}

\end{document}